\documentstyle[epsf,12pt]{article} 

\def\new{\newcommand}
\def\renew{\renewcommand}
\new{\nrao}{naive realism about operators}
\new{\Sc}{Schr\"{o}dinger}   
\new{\se}{Schr\"odinger's equation} 
\new{\BM}{Bohmian mechanics}
\new{\qf}{quantum formalism}  
\new{\qm}{quantum mechanics}
\new{\qt}{quantum theory}
\new{\wf}{wave function}       
\new{\ewf}{effective wave function}
\new{\cwf}{conditional wave function}
\new{\qe}{quantum equilibrium} 
\new{\C}{{\sf I\!\!\!C}}
\renew{\Delta}{{{\bold \nabla}}^{\;2}}
\new{\be}{\begin{equation}}
\new{\ee}{\end{equation}}
\new{\eq}[1]{(\ref{#1})}
\new{\bold}[1]{\mbox{\boldmath$#1$\unboldmath}} 
\renew{\a}{\alpha}      
\new{\ot}{\otimes}                   
\new{\psia}{\psi_{\a}}             
\renew{\H}{\mbox{${\cal H}$}}      
\renew{\P}{\mbox{${\rm I\!P}$}}
\new{\R}{\mbox{${\rm I\!R}$}}             
\new{\E}{\mbox{${\cal E}$}}
\renew{\r}{\rho}
\new{\psisq}{|\psi|^2}
\new{\born}{\r=|\psi|^2}    
\begin{document}
\title{Bohmian Mechanics  and the Meaning of the\\ Wave Function}
\author{ D. D\"{u}rr\\
Mathematisches Institut der Universit\"{a}t M\"{u}nchen\\
Theresienstra{\ss}e 39, 80333 M\"{u}nchen, Germany
\and
S. Goldstein\\
Department of Mathematics, Rutgers University\\ 
New Brunswick, NJ 08903, USA
\and
N. Zangh\`{\i}\\
Istituto di Fisica dell'Universit\`a di Genova, INFN \\ 
via Dodecaneso 33, 16146 Genova, Italy} 
\date{December 12, 1995}
\maketitle 
\openup.84\jot

\section{Introduction}\label{intro} 
Despite its extraordinary predictive successes, quantum mechanics has,
since its inception some seventy years ago, been plagued by conceptual
difficulties. Few physicists have done more than Abner Shimony to remind
us of this somewhat unpleasant fact. The most
commonly cited of these difficulties is the measurement problem,
or, what amounts to more or less the same thing, the paradox of
Schr\"odinger's cat.  Indeed, for many physicists the  measurement problem
is not merely one of the conceptual difficulties of quantum mechanics; it is
{\it the \/} conceptual difficulty.

While we have a good deal of sympathy for this view, we believe  that the
measurement problem is merely a manifestation of a more fundamental
conceptual inadequacy: It is far from clear just what it is that quantum
mechanics is about. What, in fact, does quantum mechanics describe? Many
physicists pay lip service to the Copenhagen interpretation, and in
particular to the notion that quantum mechanics is about results of
measurement. But hardly anybody truly believes this anymore---and it is hard
to believe anyone really ever did.  It seems clear now to any student of the
subject that quantum mechanics is fundamentally about atoms and electrons,
quarks and strings, and not primarily about those particular macroscopic
regularities associated with what we call measurements.

It is, however, generally agreed that any quantum mechanical
system---whether of atoms or electrons or quarks or strings---is
completely described by its wave function, so that it is also widely
accepted that quantum mechanics is fundamentally about the
behavior of wave functions.  The measurement problem provides a dramatic
demonstration of the severe difficulty one faces in attempting to maintain
this view.  

We have argued elsewhere \cite{DGZ92} that if one focuses directly on the
question as to what quantum mechanics is about, one is naturally led to the
view that quantum mechanics is fundamentally about the behavior of
particles, described by their positions---or fields, described by field
configurations, or strings, described by string configurations---and only
secondarily about the behavior of wave functions.  We are led to the view
that the wave function does not in fact provide a complete description or
representation of a quantum system and that the complete description of the
system is provided by the configuration $Q$ defined by the positions ${\bf
Q\/}_k$ of its particles together with its wave function. We are led in
fact, for a nonrelativistic system of particles, to Bohmian mechanics, for
which the {\it state\/} of the system is $(Q,\psi)$, 
which evolves according to the equations of motion
\begin{equation}\label{be}
\frac{dQ}{dt}=\mbox{Im}\frac{\nabla\psi}{\psi}(Q),
\end{equation}
where $\nabla$ is a configuration-space gradient, and
\begin{equation}\label{se}
i\frac{\partial\psi}{\partial t}=H\psi,
\end{equation}
where $H$ is the Schr\"odinger Hamiltonian.  This deterministic theory of
particles in motion, with trivial modifications to deal with spin,
completely accounts for all the phenomena of nonrelativistic quantum
mechanics, from spectral lines to interference effects, and it does so in a
completely ordinary manner. It was first presented, in a somewhat more
complicated but completely equivalent form, by David Bohm more than forty
years ago \cite{Bohm52}. Moreover, a preliminary version of this theory was
presented by de Broglie almost at the inception of quantum mechanics.  Its
principal advocate for the past three decades was John Bell \cite{Bell}.

We will here outline how Bohmian mechanics works: how it deals with various
issues in the foundations of quantum mechanics and how it is related to the
usual quantum formalism. We will then turn to some objections to Bohmian
mechanics, raised perhaps most forcefully by Abner Shimony. These
objections will lead us to our main concern: a more careful consideration
of the meaning of the wave function in quantum mechanics as suggested by a
Bohmian perspective. We wish now to emphasize, however, that a grasp of the
meaning of the wave function as a representation of a quantum system is
crucial to achieving a genuine understanding of quantum mechanics from any
perspective.

\section{The  Measurement Problem}\label{mp} 

Suppose that we analyze the process of  measurement in quantum mechanical
terms. The after-measurement wave function for system and apparatus arising
from Schr\"odinger's equation for the composite system typically involves a
superposition over terms corresponding to what we would like to regard as
the various possible results of the measurement---e.g., different pointer
orientations. Since it seems rather important that the actual result of the
measurement be a part of the description of the after-measurement situation,
it is difficult to see how this wave function could be the complete
description of this situation.  By contrast, with a theory or interpretation
like Bohmian mechanics in which the description of the after-measurement
situation includes, in addition to the wave function, at least the values of
the variables that register the result, the measurement problem vanishes.

The remaining problem of then justifying the use of the ``collapsed'' wave
function---corresponding to the actual result---in place of the original
one is often confused with the measurement problem. The justification for
this replacement is nowadays frequently expressed in terms of decoherence.
One of the best descriptions of the mechanisms of decoherence, though not
the word itself, was given by Bohm in 1952 \cite{Bohm52} as part of his
explanation of why from the perspective of Bohmian mechanics this
replacement is justified as a practical matter.  (See also \cite{DGZ92}.)

Moreover, if we focus on what should be regarded as the wave function, not
of the composite of system and apparatus, which strictly speaking remains a
superposition if the composite is treated as closed during the measurement
process, but of the system itself, we find that for Bohmian mechanics this
does indeed collapse, precisely as described by the quantum formalism. The
key element here is the notion of the conditional wave function of a
subsystem of a larger system, described briefly in section \ref{wfs} below,
and discussed in some detail, together with the related notion of the
effective wave function, in \cite{DGZ92}.

\section{The Two-Slit Experiment}\label{ts} 

Bohmian mechanics resolves the dilemma of the appearance, in one and the
same phenomenon, of both particle and wave properties in a rather trivial
manner: Bohmian mechanics is a theory of motion describing  a particle (or
particles) guided by a wave. For example, in Figure 1 we have a family of
Bohmian trajectories for the two-slit experiment. Notice that while each
trajectory passes through but one of the slits, the wave passes through both,
and the interference profile that therefore develops in the wave generates a
similar pattern in the trajectories guided by this wave. 

\begin{figure}[h]\label{tse}  
\begin{center} 
\leavevmode 
\epsfxsize=12cm
\epsfysize=4cm  
\epsffile{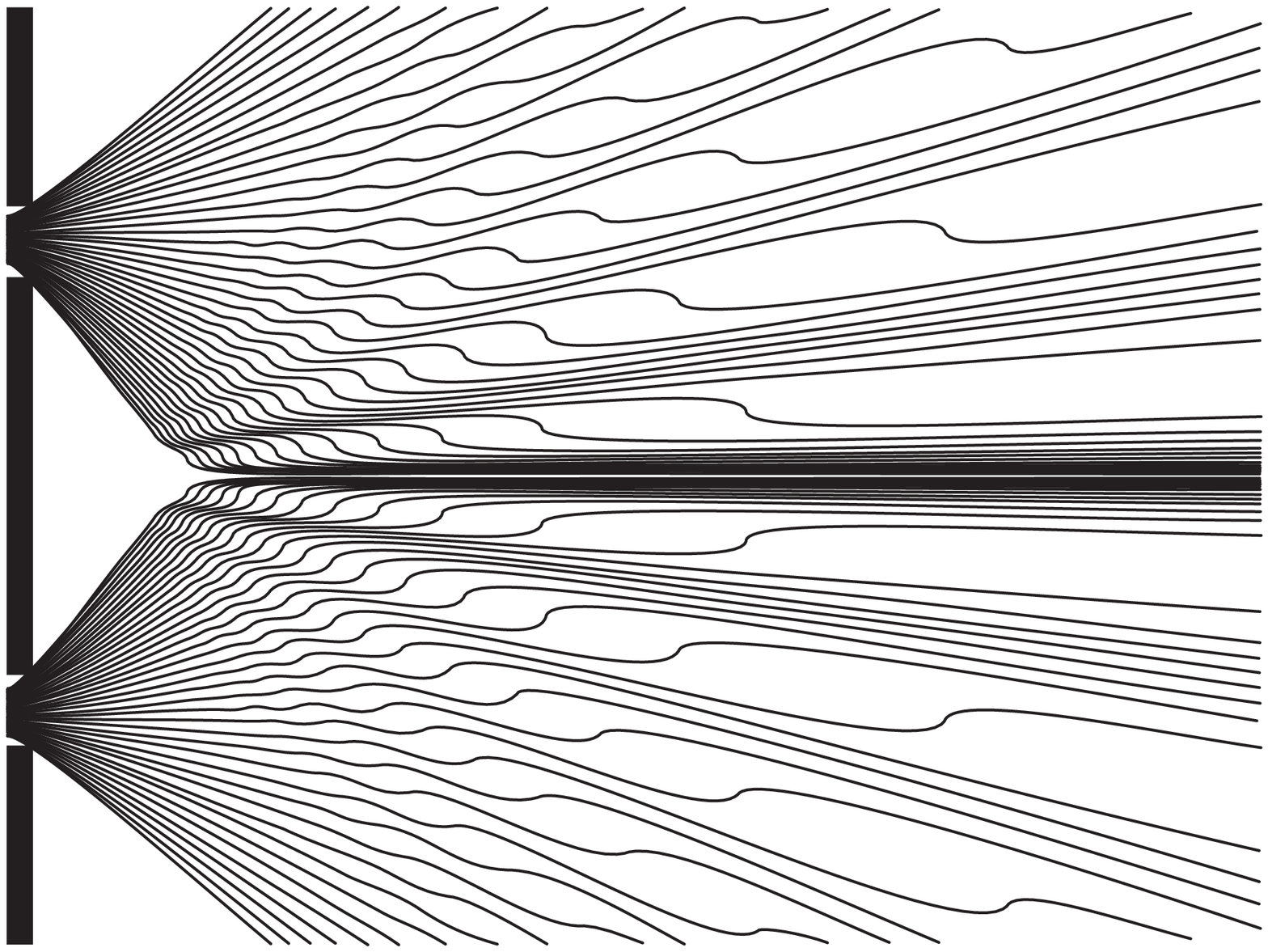} 
\end{center} 
\caption{An ensemble of trajectories for the two-slit experiment, uniform
in the slits. (Drawn by G. Bauer from [8].)}
\end{figure}

\section{The Detailed Equations and Nonlocality}
\def\pvec#1,#2{(#1_1,\dots,#1_{#2})}
\def\pd#1,#2{\frac{\partial#1}{\partial#2}}

We have given, in  (\ref{be}) and (\ref{se}),   the equations of Bohmian
mechanics in  a somewhat schematic form, without explicitly exhibiting the
parameters required for a detailed specification of the theory. Less
schematically, the equations defining Bohmian mechanics for an $N$-particle
universe of spinless particles with masses $m_k$ interacting via the
potential energy function $V=V(q)$ are   
\begin{equation}\label{dbe}
\frac{d{\bf
Q}_k}{dt}={\bf v}_k^\psi\pvec{{\bf
Q}},N\equiv\frac\hbar{m_k}\mbox{Im}\frac{\nabla_{{\bf
q}_k}\psi}{\psi}\pvec{\bf Q},N
\end{equation}
\begin{equation}\label{dse}
 i\hbar\pd \psi,t=-\sum_{k=1}^N{\frac{\hbar^2}{2m_k}\nabla_{{\bf
q}_k}^2\psi}+V\psi
\end{equation}
We have given these more detailed equations here in order to emphasize two
points. First of all, Bohmian mechanics is manifestly nonlocal, since the
velocity of any one of the particles, as expressed in (\ref{dbe}), will
typically depend upon the positions of the other particles. Thus does
Bohmian mechanics make manifest  that most dramatic effect of quantum
theory, quantum nonlocality, that Abner Shimony has so effectively
expounded. As John Bell \cite[page 115]{Bell} has stressed,

\begin{quotation}\openup-2\jot
That the guiding wave, in the general case, propagates not in
ordinary three-space but in a multidimen\-sional-configuration space is the
origin of the notorious `nonlocality' of quantum mechanics. It is a merit
of the de Broglie-Bohm version to bring this out so explicitly that it
cannot be ignored. ({\it Bell\/} 1980)
\end{quotation}

Second, we wish to emphasize that a Bohmian universe with potential $V$ is
completely specified by these two equations. Whatever is true of such a
universe must be so merely by virtue of these equations, without the
addition of further postulates such as, for example, an axiom governing the
results of momentum measurements.   And thus it is with regard to probability.

\section{Probability}\label{p}

According to the quantum formalism, the probability density for finding a
system  whose wave function is $\psi$ at the configuration $q$ is
${|\psi(q)|}^2$. To the extent that the results of measurement are
registered configurationally, at least potentially, it follows that the
predictions of Bohmian mechanics for the results of measurement  must agree
with those of orthodox quantum theory (assuming the same Schr\"odinger
equation for both) provided that it is somehow true for Bohmian mechanics
that configurations are random, with distribution given by the {\it quantum
equilibrium\/} distribution  ${|\psi|}^2$. Now the status and justification
of this quantum equilibrium hypothesis is a rather delicate matter, one that
we have explored in considerable detail elsewhere \cite{DGZ92}. We would like
to mention here but a few relevant points.

It is nowadays a rather familiar fact that dynamical systems quite
generally give   rise to behavior of a statistical character, with the
statistics given by the (or a) stationary probability distribution for the
dynamics. So it is with Bohmian mechanics, except that for the Bohmian
system stationarity is not quite the right concept, and it is rather the
notion of {\it equivariance\/} that is relevant. We say that a probability
distribution $\rho^\psi$ on configuration space, depending upon the wave
function $\psi$, is equivariant if 
\begin{equation}
\left(\rho^\psi\right)_t=\rho^{\psi_t}
\end{equation}
where the dependence on $t$ on the right arises from Schr\"odinger's
equation and on the left from the evolution on probability densities arising
from the flow (\ref{be}). Thus equivariance expresses the mutual compatibility,
relative to  $\rho^\psi$, of the Schr\"odinger evolution (\ref{se}) and
the Bohmian motion (\ref{be}). 

Now the crucial point is that $\rho^\psi={|\psi|}^2$ is equivariant, a more
or less immediate consequence of the elementary fact that the quantum
probability current $J^\psi=\rho^\psi v^\psi$, where $v^\psi$ is the r.h.s.
of (\ref{be}).  We thus have that 
\begin{center}{\it $\rho_{t_0}(q) = |\psi_{t_0}(q)|^2 $ at some time $t_0$ 
$\Longrightarrow$\\ $\rho_t(q) = |\psi_t(q)|^2$ for all $t$  }\end{center}

It is perhaps helpful, in trying to understand the status in Bohmian
mechanics of the quantum equilibrium distribution, to think of 
\begin{equation}
\mbox{\it quantum equilibrium\/}\qquad \rho = |\psi|^2
\end{equation}
as roughly analogous to (classical)
\begin{equation}
\mbox{\it thermodynamic equilibrium\/}\qquad 
\rho\sim e^{-\beta H_{\mbox{\footnotesize class}}} 
\end{equation}

\section{Operators as Observables}\label{oo}

It would appear that inasmuch as orthodox quantum theory supplies us with
probabilities not merely for positions but for a huge class of quantum
observables, it is a much richer theory than Bohmian mechanics, which seems
exclusively concerned with positions. Appearances would, however, be
misleading. In this regard, as with so much else in the foundations of
quantum mechanics, the crucial observation has been made by Bell \cite[page
166]{Bell}:

\begin{quotation}\openup-2\jot\noindent\dots in physics the only
observations we must consider are position observations, if only the
positions of instrument pointers. It is a great merit of the de
Broglie-Bohm picture to force us to consider this fact. If you make axioms,
rather than definitions and theorems, about the `measurement' of anything
else, then you commit redundancy and risk inconsistency. ({\it Bell\/}
1982) 
\end{quotation}

Now when it comes to ``definitions and theorems'' we find \cite{DDGZ} that
Bohmian mechanics leads to a natural association between an experiment
${\cal E}$ and a ``generalized observable'' defined by a
Positive-Operator-Valued measure or POV \cite{Davies} $O(dz)$ (on the value
space for the result of the experiment)
\begin{equation}\label{pov} {\cal E}  \mapsto O(dz)\end{equation}
This association is such that the probability distribution $\mu_Z^\psi(dz)$
of the result $Z$ of the experiment, when performed upon a system with wave
function $\psi$, is given by
\begin{equation}\label{distr}
\mu_Z^\psi(dz)=\langle\psi,O(dz)\psi\rangle
\end{equation}

The simplest instance of a POV is a standard quantum observable,
corresponding to  a self-adjoint operator $A$ on the Hilbert space of
``states.''  We find that more or less every ``measurement-like''
experiment  ${\cal M}$ is associated with this special kind of POV
\begin{equation}\label{sa}
{\cal E}={\cal M}\mapsto A 
\end{equation}
and we thus recover the familiar measurement axiom that the statistics for
the result of the ``measurement of the observable $A$'' are given by the
spectral measure for $A$ relative to $\psi$. 

Moreover, the conclusion (\ref{pov}) is basically an immediate consequence
of the very meaning of an experiment from a Bohmian perspective: a
coupling of system to apparatus leading after a time $t$ to a result
$Z=F(Q_t)$ that is  a function of the final configuration $Q_t$ of system
and apparatus, e.g., the orientation of a pointer on the apparatus. It
follows that the experiment ${\cal E}$ defines the following sequence of maps
\begin{eqnarray*}\psi \mapsto
\Psi = \psi \otimes \Phi_0 \mapsto \Psi_t=e^{-iHt}\Psi
\mapsto \mu(dq) =
\Psi_t^{*} \Psi_t dq \mapsto {\mu}_Z(dz) := \mu ( F^{-1}(dz)),\end{eqnarray*}
from the initial wave function of the system, to the initial wave function
of system and apparatus, to  the final wave function of system and
apparatus, to the distribution of the final configuration of the  system
and apparatus, to the distribution of the result. Thus the map
\begin{equation}\label{bl}
\psi\mapsto\mu_Z^\psi
\end{equation}
is bilinear, since each of the maps in the sequence is linear except for
the map to the quantum equilibrium distribution, which is bilinear.  
Such a bilinear map (\ref{bl}) is equivalent to a POV.

\section{The Wave Function of a Subsystem}\label{wfs} 
The existence of configurations in Bohmian mechanics as part of the reality
leads, naturally enough, to many advantages over the orthodox view that the
wave function provides us with a complete description of a physical
system. One of these advantages is that it permits a clear and natural
notion for the wave function of a subsystem of a larger system, say the
universe, a notion that from an orthodox perspective is surprisingly
problematical. Indeed, if we insist that the wave function is everything, it
is not at all clear what, in fact, is to be meant by the wave function of
anything that is directly of interest.

Let $\Psi_t$ be the wave function of the universe (at time t), and
decompose the configuration of the universe $Q=(X,Y)$ into the
configuration $X$ of the system of interest, the $x$-system, and the
configuration $Y$ of the environment of the $x$-system, i.e., the
configuration of the rest of the universe. Then we define the {\it
conditional wave function\/} of the $x$-system at time $t$ by

\begin{equation}\label{cwf}
\psi_t(x)=\Psi_t(x,Y).
\end{equation}
This turns out to be just the right notion for the wave function of a
subsystem. Moreover, under appropriate conditions it satisfies
Schr\"odinger's equation for the  $x$-system and is indeed the {\it
effective wave function\/} of the  $x$-system. See \cite{DGZ92} for details. 

\section{The Role of the Wave Function} 
In this brief section we wish to emphasize one simple point about the
structure of Bohmian mechanics: that this theory of motion is a {\it
first-order theory\/}, in which it is the first derivative of the
configuration with respect to time, rather than the second, that the theory
directly specifies. And the role of the wave function in this theory,
expressed by the association
\begin{equation}\label{vpsi}
\Psi\mapsto v^\Psi,
\end{equation}
is to generate the vector field, given by the right hand side of
(\ref{dbe}), that defines the motion.

\section{Quantum Cosmology}\label{qc}

Quantum cosmology is an embarrassment for the orthodox interpretation of
quantum mechanics as concerning merely the results of measurement---by an
external observer. When it is the entire universe with which we are
concerned, there would seem to be no room for such an observer. For Bohmian
mechanics, by contrast, there is no difficulty whatsoever on this score.

Moreover, there is another difficulty in quantum cosmology that Bohmian
mechanics greatly alleviates. The wave function $\Psi$ of the universe, as
given by a solution of the  Wheeler-de Witt equation, which we may
schematically represent by
\begin{equation}
{\cal H}\Psi=0,
\end{equation}
is stationary, and one must thus address the problem of accounting for the
emergence of change in a universe whose wave function is timeless. Now for
Bohmian mechanics we have no such difficulty, since a timeless wave
function can easily generate a nontrivial dynamics.
	
It is true that for Bohmian mechanics as defined by (\ref{be}) and
(\ref{se}), the ground state wave function, because it may be taken to be
real, generates the trivial motion. However, this will not  be true for
the generic stationary state. More important, when we contemplate a Bohmian
mechanics for quantum cosmology, we do not have in mind any particular
form for the right hand side of (\ref{be}) and in particular it need not be
the case for a Bohmian mechanics understood in this general sense---what we
have called elsewhere a Bohmian theory \cite{bmqt}---that a ground state wave
function  generates the trivial motion.

\section{Important Criticisms}\label{ic}

The most serious problem with Bohmian mechanics, (\ref{dbe}) and (\ref{dse}),
is that it manifestly fails to be Lorentz invariant. We have little to say
about this very important issue here, beyond reminding our readers that
nonlocality is an established fact that poses a challenge, not just for a
Bohmian theory, but for any precise version of quantum theory. (For some
steps in the direction of the formulation of a Lorentz invariant Bohmian
theory, as well as some reflections on the problem of Lorentz invariance,
see \cite{epr}.)  

We wish to focus here upon  two  objections. First of all, as
has been emphasized by  Abner Shimony, Bohmian mechanics violates the
action-reaction principle that is central to all of modern physics, both
classical and (non-Bohmian) quantum: There is no back action of the
configuration upon the wave function, which evolves, autonomously,
according to  Schr\"odinger's equation,  
\begin{equation}
\Psi\longrightarrow Q \qquad\mbox{but}\qquad
Q\stackrel{\mbox{\footnotesize not}} \longrightarrow\Psi
\end{equation}
Second of all, the wave function 
\begin{equation}
\Psi=\Psi\pvec{\bf q},N,
\end{equation}
which is part of the state description of---and hence presumably part of the
reality comprising---a Bohmian universe, is not the usual sort of physical
field  on physical space to which we are accustomed, 
but a field on the abstract space of all possible configurations, a space of
enormous dimension, a space constructed, it would seem, by physicists as a
matter of convenience.

\section{Some Responses}\label{sr}

Perhaps the simplest response we might make is: So what? That's
just the way it is, the way world works. Bohmian mechanics is well
defined, and who are we---as Bohr once asked of Einstein, though for a
slightly different purpose---to tell God what kinds of structures to use in
creating a world. 

We might also respond that in classical physics the
action-reaction principle is more or less an expression of conservation of
momentum, which is itself an expression of Galilean invariance (more
precisely, of translation invariance). Most physicists would also say the
same thing concerning quantum mechanics. However, in Bohmian mechanics,
because it is a first-order theory, we are able to achieve Galilean
invariance despite the no-back-action. In other words, Bohmian mechanics is
based on a fundamentally different sort of structure than   classical
mechanics, one that does not require the action-reaction principle to
achieve the desired underlying symmetry.

It might also be mentioned that the wave function of a subsystem, the
conditional wave function  (\ref{cwf}), will in general be affected by the 
configuration, via its dependence  upon the configuration
of the environment. 

However, we think that these responses  don't go far enough. We think
that the problems just mentioned suggest that we give more careful
consideration to just what sort of entity the wave function is and how it
should be regarded. Indeed, we think that both of the above objections
point in the same direction for an answer: to the question of the meaning
of the wave function.

\section{The Wave Function as  LAW}\label{wfl}
\def\ovec#1,#2{#1_1,\dots,#1_{#2}}

We propose that the reason, on the universal level, that there is no action
of configurations upon wave functions, as there seems to be between all
other elements of physical reality, is that the wave function of the
universe is not an element of physical reality. We propose that the wave
function belongs to an altogether different category of existence than that
of substantive physical entities, and that its existence is nomological rather
than material. We propose, in other words, that the wave function is a
component of physical law rather than of the reality described by the law.

We note in this regard that nobody objects to classical mechanics because it
involves a Hamiltonian $H_{\mbox{\footnotesize class}}(\ovec {{\bf
q}},N,\ovec {{\bf p}},N )\equiv H_{\mbox{\footnotesize class}}(\xi)$ that is
a function on a space, the phase space, that is of greater dimension and
even more abstract than configuration space. This is because we think of the
state in classical mechanics as given by the $q$'s and $p$'s, and we regard
the Hamiltonian as the generator of the evolution of the state---i.e., as
part of the law---and not as an object in whose behavior we are directly
interested.

To pursue this analogy, between the wave function and the classical
Hamiltonian, a bit further, let's compare 
\begin{equation}
H_{\mbox{\footnotesize class}}\longleftrightarrow\log\Psi
\end{equation}
and note that both of these generate motions in pretty much the same way
\begin{equation}
\frac{d\xi}{dt}=\mbox{Der}H_{\mbox{\footnotesize
class}}\longleftrightarrow\frac{dQ}{dt}=  \mbox{Der}(\log\Psi), 
\end{equation}
with Der a derivation. Moreover, when we proceed to the level of
statistical mechanics, we find statistics of the more or less the same form
\begin{equation}
\rho_{\mbox{\footnotesize class}}\sim e^{\mbox{\footnotesize
const.}H_{\mbox{\footnotesize 
class}}}\longleftrightarrow \rho_{\mbox{\footnotesize
quant}}\sim|e^{\mbox{\footnotesize const.}\log\Psi}|,
\end{equation}
(with the constant on the right equal to 2).

Now we do not think that this analogy should be taken too seriously or too
literally; it's not a particularly good analogy---but it's better than it
has any right to be. It does, however, have the virtue that it stimulates a
new direction of thought concerning the meaning of the wave function, and
that is a great virtue indeed. 

Perhaps the most serious weakness in the analogy is that, unlike
$H_{\mbox{\footnotesize class}}$, $\psi=\psi_t$ is time-dependent, and
indeed is a 
solution of what we regard as the {\it fundamental} equation of {\it
motion} for $\psi$, \begin{equation} i\frac{\partial\psi}{\partial
t}=H\psi.\end{equation} 
Moreover, for a particular choice of classical theory, with specified
interactions, 
$H_{\mbox{\footnotesize class}}$ is fixed; it is not free, not something to be
chosen as an initial condition, like $\psi$. 

But think now again of the Wheeler-de Witt equation for {\it the\/} wave
function of the universe. This fundamental wave function $\Psi$, the universal
wave function, is static, stationary, and, in the view of many physicists,
unique.  The fundamental equation for  $\Psi$  

\begin{equation}\label{HPsi0}
{\cal H}\Psi=0
\end{equation}
or more generally
\begin{equation}\label{HPsiE}
{\cal H}\Psi=E\Psi			
\end{equation}
should be regarded as a sort of generalized Laplace equation that selects
the central element $\Psi$ of the law of motion
\begin{equation}\label{v}
dQ/dt=v^{\Psi}(Q),		
\end{equation}
the object that generates the vector field $v^{\Psi}$ defining the
motion. Here $Q$ is rather general---not  merely particle
positions, and certainly including the configuration of the gravitational
field. Moreover, the form of  $v^{\Psi}$ should arise from the mathematical
and geometrical character of the structure defined by $Q$, and should not
be conceived of as being of any particular a priori form, such as given in
the r.h.s. of (\ref{be}).    

The equation (\ref{v}) is now the fundamental equation of motion, with
$\Psi$ the (natural) solution to the ``Laplace equation,'' which defines
the law of motion (\ref{v}) through the selection of $\Psi$. We may regard
this selection as analogous to that of the Coulomb interaction via the
equation $\Delta\phi=\delta$. (Note also that $H_{\mbox{\footnotesize
class}}$ for the Coulomb interaction satisfies something much like Poisson's
equation on phase space, $ (\Delta_p+\Delta_q)H_{\mbox{\footnotesize
class}}=\mbox{const}+\sum\delta$.) In particular $\Psi$, and hence
(\ref{v}), does not explicitly depend upon time $t$---since there is no $t$
in (\ref{HPsi0}) or (\ref{HPsiE}).

\section{The Schr\"odinger Evolution as Phenomenological}\label{sep}

We wish to stress  that we are now exploring the possibility that
the time-dependent Schr\"odinger equation is not fundamental. We must thus
address the question, not of how change is at all possible in a theory with
a change-less wave function---since this is trivial when, in addition to the
wave function, there is the configuration $Q$ whose very motion it is the
role of the wave function to specify---but rather why we should arrive, as
we do, at a picture with time-dependent Schr\"odinger wave function when we
start with a theory with a fixed timeless wave function that knows nothing
of the time-dependent Schr\"odinger equation.  

Now we already know that for Bohmian mechanics the Schr\"odinger evolution
is hereditary, so that if the  universal wave function $\Psi$ satisfies
Schr\"odinger's equation then subsystems will (in the usual situations and
under the usual assumptions, see \cite{DGZ92}) have their own wave functions,
nontrivially evolving according to their own Schr\"odinger evolutions. Since
a wave function satisfying (\ref{HPsiE}) {\it does\/} define a solution to
Schr\"odinger's (albeit a very special one), we should perhaps expect to
find subsystems behaving as just described even for a theory in which the
time-dependent Schr\"odinger evolution is not fundamental.

However, since it may not be clear how a stationary wave function {\it
could\/} yield an evolution rich enough to generate genuinely evolving
subsystem wave functions,\footnote{Note that in the usual measurement theory
picture, it is the {\it motion\/} of the composite system wave function that
appears to be directly responsible for the motion of the ``collapsed'' wave
function.}  we wish to give a very simple example in which this occurs, as
well as to tentatively propose a more general analysis.   

Suppose that the configuration of the universe has a decomposition of the
form 
\begin{equation}
q=(x,y)
\end{equation}
\begin{equation}Q(t)=(X(t),Y(t)),\end{equation}
where $X$ describes the degrees of freedom with which we are somehow most
directly concerned and $Y$ describes the remaining degrees of freedom. For
example, $X$ might be the configuration of all the degrees of freedom
governed by standard quantum field theory, describing the fermionic matter
fields as well as the bosonic force fields, while $Y$ refers to the
gravitational degrees of freedom. We wish to focus on the conditional wave
function  

\begin{equation}\psi_t(x)=\Psi(x,Y(t))
\end{equation}
for the $x$-system and to ask whether $\psi_t(x)$ could be---and might,
under suitable conditions, be expected to be---a solution to Schr\"odinger's
equation for the $x$-system.

First, the simple example: Suppose our universe consists merely of two
particles, with configurations $x$ and $y$ respectively, moving in a
1-dimensional space. Suppose further that the particles are noninteracting,
so that the l.h.s. of (\ref{HPsiE}) is just the free Hamiltonian
($\hbar=m_k=1$) 
\bigskip
\begin{equation}\label{Hsum}
H=-\frac12\Delta=-\frac12(\frac{\partial^2}{{\partial
x}^2}+\frac{\partial^2}{{\partial y}^2})=H_x+H_y
\end{equation}
Let 

\begin{equation}
\label{ecos}
\Psi(x,y)=e^{i(x-y)}\cos(x+y).
\end{equation}
This ``wave function of the universe'' satisfies (\ref{HPsiE}) with $E=2$
\begin{equation}H\Psi=2\Psi
\end{equation}
[This wave function  is of course best arrived at by rotating the obvious 
eigenfunction $e^{ikx}\cos{ky}$ ($k=\sqrt2$) by 45 degrees.]

It then follows immediately from (\ref{be}) that
\begin{equation}
Y(t)=y_0 - t,
\end{equation}
so that the conditional wave function  
\begin{equation}\psi_t(x)\sim e^{i(x+t)}\cos(x+y_0-t)\equiv
e^{2it}\hat\psi_t(x) 
\end{equation}
is clearly not stationary and moreover is (projectively and hence
physically) equivalent to $\hat\psi_t$, which satisfies
\begin{equation}i\frac{\partial\hat\psi}{\partial t}=H_x\hat\psi.
\end{equation}

\smallskip 
We will now present an argument suggesting that what we've just found in
the example---a time-dependent conditional wave function obeying
Schr\"odinger's equation emerging from a stationary universal wave
function---should be expected to occur much more generally. Suppose we can
write
\begin{equation}\label{asum} 
\Psi(x,y)\simeq\sum_\alpha \psi_t^\alpha(x)\phi_t^\alpha(y)
\end{equation}
where for each $t$, $\phi_t^\alpha(y)$ is a ``narrow wave packet,''
centered around $y_t^{\alpha}\ [\neq y_t^{{\alpha}^\prime}]$. 
Suppose that the time-dependence in (\ref{asum}) is such that 
$\phi_t^\alpha(y)$ ``follows'' $Y(t)$, i.e., that
$Y(t)\approx y_t^{\alpha}$ for all $t$, where $\alpha$ is such that
$Y(0)\approx y_0^{\alpha}$. 
It then follows from  (\ref{asum}) that for the conditional wave function  of
the $x$-system we have that $\psi_t(x)\approx \psi_t^\alpha(x)$.

Now we know what kind of time-dependence is such that $\phi_t^\alpha(y)$
keeps up with $Y(t)$. This occurs when $\phi_t^\alpha(y)$ is a solution of
Schr\"odinger's equation (with Hamiltonian $H_y$). Since $\Psi$ itself has
no time-dependence in it, a natural way to arrive at  (\ref{asum}) is to
consider a single decomposition of the form (\ref{asum}), involving narrow
and approximately disjoint $y$-wave packets, say for $t=0$, and write
\begin{eqnarray*}
\Psi\sim
e^{-iEt}\Psi=e^{-iHt}\Psi=&e^{-i(H_x+H_y)t}\displaystyle\sum_\alpha
\psi_0^\alpha(x)\phi_0^\alpha(y)\\=&\displaystyle\sum_\alpha
\left(e^{-iH_xt}\psi_0^\alpha(x)\right)
\left(e^{-iH_yt}\phi_0^\alpha(y)\right)\\\equiv&\displaystyle 
\sum_\alpha\psi_t^\alpha(x)\phi_t^\alpha(y),
\end{eqnarray*}
from which we see that $i\frac{\partial\psi_t^\alpha}{\partial
t}=H_x\psi_t^\alpha$.\footnote{More generally, we might have considered
$e^{-i\gamma Et}\Psi$, but our desire that $y_t^{\alpha}\approx Y(t)$ leads
to the choice $\gamma=1$.} Now if, for example, we are dealing here with
the semi-classical regime for the $y$-system, an initial collection of
narrow and approximately disjoint wave packets $\phi_0^\alpha(y)$ should
remain so under their evolution.  Then  the conditional wave
function of the $x$-system will approximately satisfy
\[
i\frac{\partial\psi}{\partial t}=H_x\psi.
\]
It is perhaps worth noting that if $Y$ describes the gravitational degrees
of freedom, we might imagine that the evolution $Y(t)\approx y_t^{\alpha}$
describes the expansion of the universe.

We thus see how Schr\"odinger's (time-dependent) equation might indeed
rather generally arise as a phenomenological equation that emerges when we
look for a description of the behavior of subsystems of a universe governed
by a timeless universal wave function that knows nothing about
Schr\"odinger's equation.  

\section{Overview}\label{o}
We wish to underline the transitions in quantum ontology implied by our
discussion, proceeding from what is arguably the ontology of Orthodox Quantum
Theory, to that of Orthodox Bohmian Mechanics, and finally to the ontology
of the 
Universal Bohmian Theory upon which we have just focused:  

\begin{center}OQT\hfil OBM\hfil UBT
%\\\mbox{}
%\\\mbox{}
%\\\mbox{}
\\$\Psi\hfil(\Psi,Q)\hfil Q$\end{center}

In conclusion, we note that Bohmian mechanics is profoundly unromantic. It
tends to be a counterexample to lots of seductive notions about quantum
mechanics, for example:  
\begin{itemize}
\item many-worlds
\item observer-created reality
\item noncommutative epistemology
\item quantum logic
\end{itemize}
There is, however, one element of quantum peculiarity that Bohmian
mechanics is normally regarded as retaining and amplifying. Bell \cite[page
128]{Bell} has said that
\begin{quotation}\openup-2\jot\noindent {\it No one can understand this theory
until he is willing to think of $\Psi$ as a real objective field rather
than just a `probability amplitude.' Even though it propagates not in
$3$-space but in $3N$-space.\/} (Bell 1981)
\end{quotation}
Concerning the notion that the wave function is fundamentally, if not {\it
the\/} reality, at least a substantive part of reality, what we are
suggesting here is that Bohmian mechanics may turn out to be a
counterexample to this as well.

\section*{Acknowledgments} We are grateful to Rick Leavens for a careful
reading and valuable suggestions.  This work was supported in part by the
DFG, by NSF Grant No. DMS-9504556, and by the INFN.

\end{document}